\documentclass[aps,notitlepage,twocolumn,showpacs,preprintnumbers,floatfix,tightenlines, longbibliography, footinbib,superscriptaddress]{revtex4-1}
\usepackage{mathrsfs}
\usepackage{amssymb}
\usepackage{amsmath}
\usepackage{graphicx}
\usepackage{amsfonts,amsbsy}
\usepackage{nicefrac}
\usepackage{slashed}
\usepackage[dvipsnames]{xcolor}
\usepackage[breaklinks,colorlinks,urlcolor=blue,citecolor=citcolor,linkcolor=lcolor]{hyperref}
\usepackage{array}
\usepackage{multirow}
\usepackage{siunitx}
\usepackage[normalem]{ulem} 

\definecolor{lcolor}{rgb}{0.5,0,0}
\definecolor{citcolor}{rgb}{0,0.3,0.0}
\definecolor{ao(english)}{rgb}{0.0, 0.5, 0.0}
\definecolor{RoyalBlue}{HTML}{0071BC}

\newcommand{\mbf}{\mathbf}
\newcommand{\mrm}{\mathrm}

\newcommand{\CA}{C_{\mrm{A}}}
\newcommand{\NC}{N_\mathrm{c}}
\newcommand{\nf}{n_f} 
\newcommand{\CF}{C_{\mrm{F}}}

\newcommand{\CR}{C_{\mrm{R}}}

\newcommand{\fig}{Fig.~}

\newcommand{\qperp}{q_\perp}

\newcommand{\qhat}{\hat q}

\newcommand{\Gammael}{\Gamma_{\text{el}}}

\newcommand{\vb}{\vec}
\renewcommand{\vec}[1]{\mathrm{\mathbf{#1}}}
\newcommand{\dd}[2][]{\mathrm d^{#1}{#2}\,} 
\newcommand{\dv}[2][]{\frac{\dd{#1}}{\dd{#2}}}
\newcommand{\pdv}[2][]{\frac{\partial{#1}}{\partial{#2}}}

\newcommand{\tauR}{\tau_R}

\newcommand{\Conetwo}{\mathcal {C}^{1\leftrightarrow 2}}
\newcommand{\Ctwotwo}{\mathcal{ C}^{2\leftrightarrow 2}}

\newcommand{\Teps}{T_{\varepsilon}}

\newcommand{\xianiso}{\xi_0}

\newcommand{\tform}{t^{\mathrm{form}}}

\DeclareSIUnit\c{c}

\begin{document}

\title{Jet broadening and radiation in the early anisotropic plasma in heavy-ion collisions}
\author{Alois Altenburger}
\affiliation{Institute for Theoretical Physics, TU Wien, Wiedner Hauptstraße 8-10, 1040 Vienna,
Austria}
\author{Kirill Boguslavski} 
\affiliation{Institute for Theoretical Physics, TU Wien, Wiedner Hauptstraße 8-10, 1040 Vienna,
Austria}
\affiliation{SUBATECH UMR 6457 (IMT Atlantique, Université de Nantes, IN2P3/CNRS), 4 rue Alfred Kastler, 44307 Nantes, France}
\author{Florian Lindenbauer} 
\email{flindenb@mit.edu}
\affiliation{Institute for Theoretical Physics, TU Wien, Wiedner Hauptstraße 8-10, 1040 Vienna,
Austria}
\affiliation{MIT Center for Theoretical Physics – a Leinweber Institute, Massachusetts Institute of Technology, Cambridge, MA 02139, USA}

\preprint{MIT-CTP/5917}

\begin{abstract}
Measuring the energy loss of energetic jet partons may provide experimental opportunities to constrain the initial nonequilibrium stages in heavy-ion collisions, which requires theoretical predictions for the jet-medium potential.
In this letter, we go beyond the usual harmonic approximation, extract this potential for the first time using QCD kinetic theory simulations, and provide a simple cutoff-independent small-distance form.
We find significant differences to thermal media at early times in momentum and position space, as well as substantial angular dependence. 
Applying our results to the gluon splitting rates using a novel computational method reveals tensions in the current kinetic theory implementations for early times.
\end{abstract}

\maketitle

{\textbf{Introduction.}---}%
Upcoming high-luminosity upgrades to the Large Hadron Collider (LHC) and new lighter ion collision systems 
provide novel experimental opportunities, which may increase sensitivity to the initial pre-hydrodynamic stages of the QCD medium.
For instance, this could provide an experimental handle on the currently debated existence and properties of a deconfined QCD plasma in small systems.
At sufficiently high energies, such a QCD plasma can be described using an effective kinetic theory \cite{Arnold:2002zm},
which is commonly used to study the far-from-equilibrium initial stages and the approach to hydrodynamics \cite{Kurkela:2015qoa, Berges:2020fwq, Schlichting:2019abc}.

Jets provide a promising experimental tool to explore these initial stages.
Consisting of energetic partons, they traverse the plasma and lose energy by inelastic gluon emission. This process is known as \emph{jet quenching} and is considered a signature of the formation of the deconfined medium \cite{Qin:2015srf, Apolinario:2022vzg}. The emission process is triggered by elastic collisions with the plasma partons, which bring the jet parton slightly off shell, opening up the phase space for inelastic gluon emissions. To quantify this energy loss, several formalisms have been developed, which depend on the generalized interaction potential or \emph{dipole cross section} $C(\vb b)$ \cite{Zakharov:1996fv, Zakharov:1997uu, Zakharov:1998sv, Baier:1996kr, Baier:1996sk, Baier:2000mf, Gyulassy:1999zd, Gyulassy:2000er, Gyulassy:2003mc, Wiedemann:2000za, Salgado:2003gb, Arnold:2002ja, Djordjevic:2008iz, Arnold:2008iy, Caron-Huot:2010qjx, Blaizot:2012fh}.
Its physical interpretation is evident from the \emph{elastic collision kernel} $C(\vb q_\perp)$, which is related via a Fourier transform and 
measures the rate of transverse momentum broadening of a parton moving through the plasma.

Previous studies of jet-medium interactions, such as momentum broadening and energy loss, focused on a thermal or hydrodynamic medium, or applied a thermal form for the collision kernel even for the nonequilibrium plasma \cite{Aurenche:2002pd, Arnold:2008vd, Caron-Huot:2008zna, Andres:2020vxs, Moore:2021jwe, Schlichting:2021idr, Andres:2023jao, Soudi:2024yfy, Caron-Huot:2010qjx, Yazdi:2022bru, Modarresi-Yazdi:2024vfh, Liu:2006ug, Liu:2006he, Armesto:2006zv, DEramo:2010wup, Zhang:2012jd, Kumar:2020wvb, He:2015pra, Schlichting:2020lef, Mehtar-Tani:2022zwf, Zhou:2024ysb}.
These medium-induced gluon emissions also play an important role in QCD kinetic theory simulations, where they enter via the gluon splitting rates $\gamma$. In these simulations, they are computed
using a simple isotropized collision kernel \cite{Aurenche:2002pd, AbraaoYork:2014hbk, Kurkela:2014tea}, which neglects anisotropic effects that may be crucial out of equilibrium.

While there exist recent efforts to generalize jet quenching formalisms to anisotropic, inhomogeneous, and flowing systems \cite{Romatschke:2004au, Romatschke:2006bb, Dumitru:2007rp, Hauksson:2021okc, Sadofyev:2021ohn, Andres:2022ndd, Barata:2022krd, 
Barata:2023qds, Kuzmin:2023hko, Barata:2024xwy}, realistic medium input is needed for an exhaustive assessment of the features of such a nonequilibrium plasma. Among the first steps, the time evolution of the jet quenching parameter $\qhat(\tau)$ has been computed in the early Glasma and kinetic theory stages \cite{Ipp:2020nfu, Ipp:2020mjc, Avramescu:2023qvv, Carrington:2016mhd, Carrington:2021dvw, Carrington:2022bnv, Boguslavski:2023alu, Boguslavski:2023jvg, Boguslavski:2024jwr}, where it characterizes the small-distance ``harmonic'' approximation of the full collision kernel. While this approximation is often employed in recent phenomenological studies \cite{Andres:2019eus, Huss:2020whe, Adhya:2024nwx}, it may miss potentially relevant nonequilibrium effects.

In this work, we go beyond the small-distance or isotropic limits and obtain the full nonequilibrium jet-medium interaction potential during the initial stages using QCD kinetic theory and the gluon splitting rates.
\newline

{\textbf{Collision kernel.---}%
The collision kernel $C(\vb q_\perp)$ can be understood as a generalization of the jet quenching parameter $\hat q$. Often used in jet quenching calculations, $\hat q$ is a single number quantifying momentum broadening
\begin{align}
\hat q=\dv[\langle p_\perp^2\rangle]{\tau}=\int \frac{\dd[2]{\vb q_\perp}}{(2\pi)^2}\,\vb q_\perp^2C(\vb q_\perp). \label{eq:qhat-definition}
\end{align} 

The collision kernel encodes the probability for an energetic parton (with momentum $\vb p$) to receive an elastic transverse momentum kick $\vb q_\perp$ in the plasma due to scattering off a plasma particle with momentum $\vb k$. It can be obtained in QCD kinetic theory from the elastic scattering rate $\Gammael$ via (see the Appendix for more details)
\begin{align}
    \label{eq:formula_Cq}
    C(\vb q_\perp)&=(2\pi)^2\frac{\dd\Gammael}{\dd[2]{\vb q_\perp}}=\frac{1}{2p\nu_a}\sum_{bcd}\int\frac{\dd[3]{\vb k}}{(2\pi)^3}\dd {q_\parallel} \\
    &\times \frac{|{\hat{\mathcal M}^{ab}_{cd}}(\vb p,\vb k;\vb p+\vb q, \vb k-\vb q)|^2}{8|\vb k||\vb k-\vb q| |\vb p+\vb q|}f_b(\vb k)(1+f_d(\vb k')), \nonumber
\end{align}
where $k'=|\vb k-\vb q|$ ($p'=|\vb p + \vb q|$) 
is the momentum of the outgoing plasma (jet) particle, and $\vb q=(\vb q_\perp, q_\parallel)$ is the momentum transfer during the elastic scattering process. Note that in this coordinate system,
$\vb q_\perp = q_\perp (\cos(\phi), \sin(\phi))$ is perpendicular to $\vb p$. For a jet traversing in the $x$-direction, $\phi = 0$ ($\phi = \pi/2$) indicates momentum transfer along (perpendicular to) the beam axis~${\hat {\vb z}}$. 
Here, $f_a(\vb p)=\frac{(2\pi)^3}{\nu_a}\frac{\dd {N_a}}{\dd[3]{\vb x}\dd[3]{\vb p}}$ is the distribution function per degree of freedom.
The matrix element $|\hat {\mathcal M}|^2=|\mathcal M|^2\delta(p+k-p'-k')$ is summed over all incoming and outgoing color and spin factors ($\nu_a$ counts the degrees of freedom of the jet particle, i.e., $\nu_g=2(\NC^2-1)$ for a gluon), and includes a delta functionenforcing energy conservation.
In the matrix elements (tabulated, e.g., in Ref.~\cite{Arnold:2002zm, Boguslavski:2023waw}), we include medium-effects by using the isotropic
\footnote{This typically employed isotropic screening prescription prevents and neglects the effect of plasma instabilities, which seem to not play a significant role during the thermalization dynamics \cite{Berges:2013eia, Berges:2013fga}.
}
hard-thermal loop resummed propagator for internal soft momenta in the limit $p\to\infty$ \cite{Boguslavski:2024kbd}, which depends on the Debye screening mass $m_D$ given by
$m_D^2=4\lambda\int\frac{\dd[3]{\vb p}}{(2\pi)^3|\vb p|}f(\vb p)$ for a gluon system and with the 't Hooft coupling $\lambda=g^2\NC$.

Since we are interested in the collision kernel $C(\vb q_\perp)$, we start
with its asymptotic behavior 
that follows from Eq.~\eqref{eq:formula_Cq}.
For large $q_\perp$, we have $f(\vb k')\to 0$ and can use the vacuum form of the matrix elements $|\mathcal M|^2\sim 1/q_\perp^4$. Then, 
$C(\vb q_\perp)$ is
proportional to the number density 
\begin{align}
    \label{eq:mathcal_N}
    \mathcal N=\int\frac{\dd[3]{\vb k}}{(2\pi)^3}\left(2\NC f_g(\vb k)+\sum_{s\in\{q,\bar q\}} f_s(\vb k)\right),
\end{align}
summed over quarks and antiquarks of all flavors.
Conversely, for $\qperp\to 0$ 
and using a sum rule \cite{Aurenche:2002pd},
the leading contribution is
$\sim 1/q_\perp^2$ with a possible dependence of the prefactor on the direction of $\vb q_\perp$,
\begin{subequations}\label{eq:limits_general}
\begin{align}
    C(\vb \qperp)\to \begin{cases}
        \frac{a(\phi)}{\qperp^2}, & q_\perp \to 0\\
        \frac{g^4\CR\mathcal N}{\qperp^4}, & q_\perp \to \infty\,,
    \end{cases}\label{eq:limits_qperp}
\end{align}
with a more detailed derivation in the Appendix. Here, $\CR$ is a color factor $\CA=\NC$ for a gluon and $\CF=(\NC^2-1)/(2\NC)$ for a quark.

An isotropic $f(|\vb p|)$
yields a simple small-$q_\perp$ form \cite{Aurenche:2002pd, Arnold:2002zm}
\begin{align}
C(q_\perp \to 0) = 
g^2C_RT_\ast\,\frac{m_D^2}{q_\perp^2(q_\perp^2+m_D^2)},
\label{eq:analytic_collision_kernel}
\end{align}
\end{subequations}
with the infrared temperature $T_\ast=2\lambda\int\dd[3]{\vb p}f(\vb p)(1+f(\vb p))/(m_D^2(2\pi)^3)$ (for a gluon plasma) and Debye mass $m_D$.
In equilibrium, $\mathcal N=(4\NC+3\nf)\zeta(3)/(2\pi^2)$,
$T_\ast = T$, and $m_D^2 \equiv (m_D^{\mathrm{eq}})^2 = {g^2T^2(2\NC+\nf)}/{6}$.
The form \eqref{eq:analytic_collision_kernel} enters the isotropic screening approximation for $\Conetwo$ in kinetic theory simulations \cite{Arnold:2002zm, Kurkela:2015qoa}. 
\newline

{\textbf{Kinetic theory}.---}%
We now compute the collision kernel \eqref{eq:formula_Cq} in an expanding plasma. For simplicity, we will only consider gluons, whose distribution $f$ results from solving
the Boltzmann equation numerically \cite{Arnold:2002zm}
\begin{align}
\pdv[f(\vb p)]{\tau}- \frac{p_z}{\tau} \pdv[f_{\vb p}]{p_z} =-\Conetwo[f(\vb p)]- \Ctwotwo[f(\vb p)] .\label{eq:boltzmann_equation}
\end{align}
Here, $\Ctwotwo$ describes elastic collisions
and $\Conetwo$ near-collinear inelastic collisions. The latter include the gluon splitting rate, where the collision kernel enters.
We use an implementation based on \cite{kurkela_2023_10409474}, with the soft-gluon exchange in $\Ctwotwo$ regulated by the isotropic hard-thermal loop self-energy as described in \cite{Boguslavski:2024kbd}.
We employ a highly occupied anisotropic initial condition introduced in \cite{Kurkela:2015qoa} and inspired by the earlier Glasma stage, 
 $   f(\vb p,\tau{=}1/Q_s)=\frac{2A\langle p_T\rangle}{\lambda \, p_\xi} \exp \Big ({-\frac{2p_\xi^2}{3\langle p_T\rangle^2}}\Big), 
    \label{eq:initial_cond}$
with $p_\xi=\sqrt{p_\perp^2+(\xianiso p_z)^2}$, $\xianiso = 10$, $A = 5.24171$ and $\langle p_T\rangle = 1.8 Q_s$ as in previous studies.
For the 't Hooft couplings $\lambda=2$ and $\lambda=10$, the initialization time becomes $\tau/\tauR\approx 0.01$ and $\tau/\tauR\approx 0.09$, where $\tauR=4\pi\eta/s/\Teps$ is the relaxation time.
It includes the specific shear viscosity $\eta/s$
and an effective temperature $\Teps$ of the thermal system corresponding to the same energy density $\varepsilon=\nu\int\dd[3]{\vb p}pf(\vb p)/(2\pi)^3$ as the nonequilibrium plasma.
For details on solving Eq.~\eqref{eq:boltzmann_equation}
we refer to \cite{AbraaoYork:2014hbk, Kurkela:2014tea, Kurkela:2015qoa, Boguslavski:2024kbd}.

We evolve the distribution function via \eqref{eq:boltzmann_equation} and calculate the collision kernel using \eqref{eq:formula_Cq}. We have checked that our implementation of $C(\vb \qperp)$ is consistent by 
recovering the thermal limits \eqref{eq:limits_general},  
and with \eqref{eq:qhat-definition}, yielding the same values for
$\hat q$ as reported in Ref.~\cite{Boguslavski:2023alu}.
As in previous works \cite{Boguslavski:2023alu, Boguslavski:2023fdm, Boguslavski:2023jvg}, we include time markers to signal specific times during the simulation: the star marker is placed where the rescaled occupancy $\lambda f$ drops below one, the circle marker where it reaches its minimum and the triangle where the pressure anisotropy is $0.5$, signaling a system relatively close to isotropy.
\newline

\begin{figure}
    \centering
    \includegraphics[width=\linewidth]{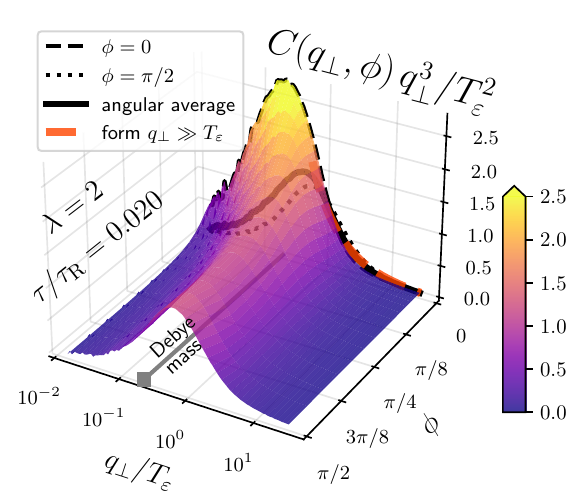}
    \includegraphics[width=\linewidth]{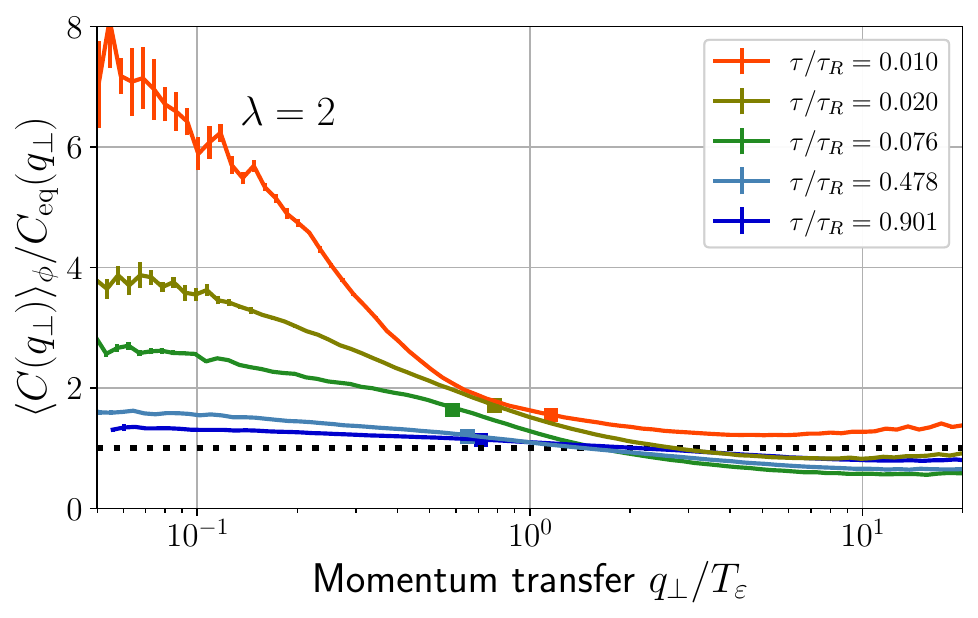}
    \caption{(Top): The integrand of $\qhat$, i.e., the collision kernel multiplied by $\qperp^3$, with angular average and smallest and largest angles.
    (Bottom): The angular averaged non-equilibrium collision kernel normalized to the numerically obtained equilibrium kernel. Squares indicate the Debye mass.
\label{fig:nonequilibrium_kernel_normalized_equilibrium}
    }
\end{figure}

{\textbf{Results for $C(\vb q_\perp)$}.---}%
We present our results for the collision kernel in Fig.~\ref{fig:nonequilibrium_kernel_normalized_equilibrium}. The top panel shows the collision kernel multiplied by $q_\perp^3$ for an early time as a function of the magnitude $q_\perp$ and the azimuthal angle $\phi$ of $\vb q_\perp$. This corresponds to the integrand of the jet quenching parameter (c.f.~Eq.~\eqref{eq:qhat-definition})
and is made dimensionless by division 
by $T_\varepsilon^2$. The Debye mass $m_D$ is marked by a square, where the kernel is peaked for a thermal system. 
We observe that, as expected at large $q_\perp$, the kernel becomes isotropic and follows the large $q_\perp$ limit \eqref{eq:limits_qperp} shown as a red dashed line. 
In contrast, 
for early times as shown here, the kernel is peaked at lower momenta $q_\perp$ and small angles $\phi$, taking its maximum at $\phi=0$, i.e., along the beam axis. This may be interpreted as an effectively emergent angle-dependent screening scale for momentum broadening with increasingly efficient broadening in the beam direction. We find qualitatively similar features for $\lambda=10$ as shown in the Appendix.

The bottom panel shows the angular averaged non-equilibrium kernel $\langle C(\vb q_\perp)\rangle_\phi = \int \frac{\dd\phi}{2\pi}C(\vb q_\perp)$ 
normalized to a thermal plasma with temperature $\Teps$. 
We find that at early times, the collision kernel is enhanced at small momenta, while reduced at large momenta. This supports the often-employed small-angle approximation for elastic scatterings \cite{Mueller:1999pi, Baier:2000sb, Hong:2010at, Blaizot:2013lga, Blaizot:2014jna, Brewer:2022vkq, Rajagopal:2024lou, Rajagopal:2025nca}. 
\newline

\begin{figure*}
    \centerline{
        \includegraphics[width=0.5\linewidth]{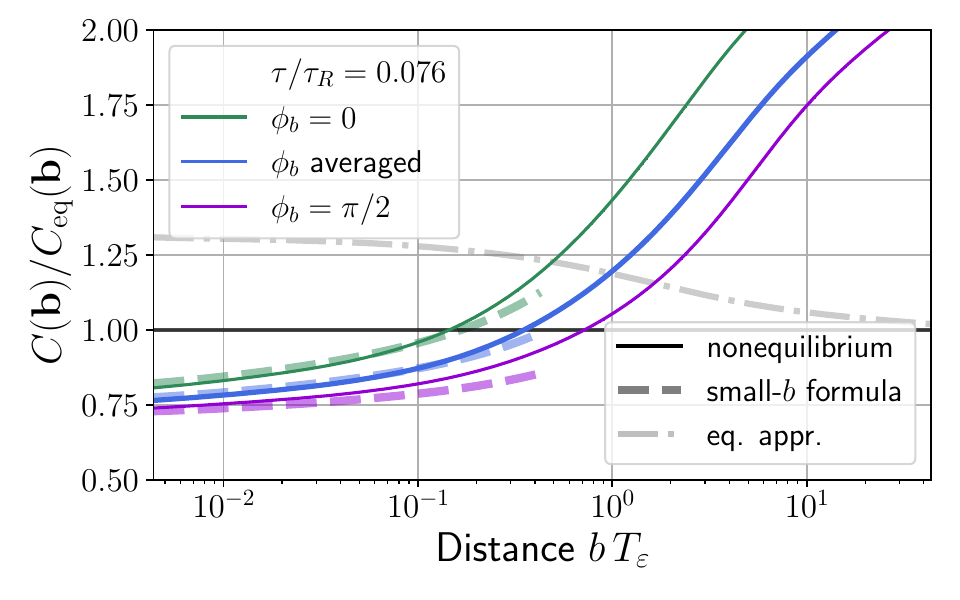}
        \includegraphics[width=0.5\linewidth]{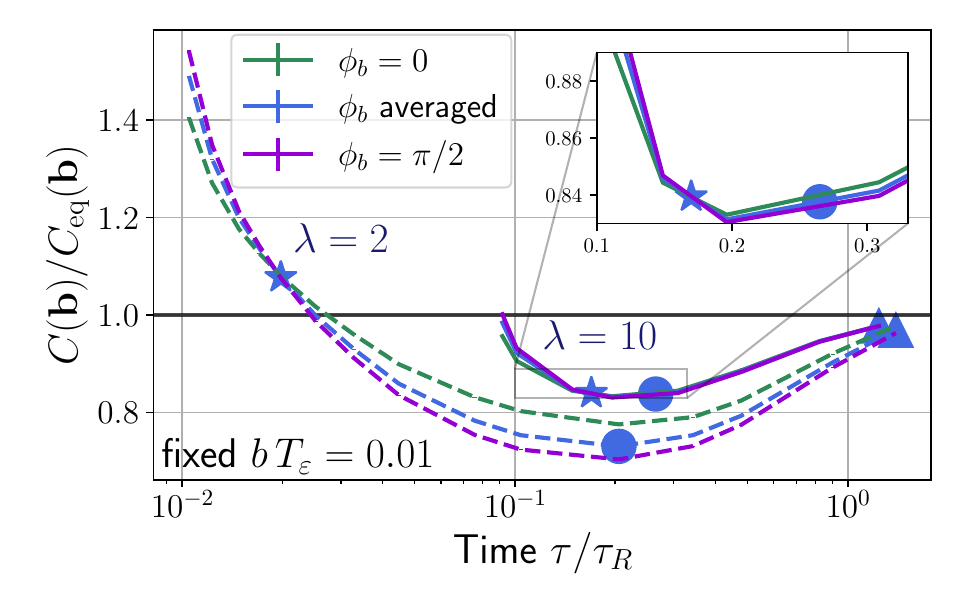}
    }
    \caption{Dipole cross section $C(\vb b)$ in impact parameter space normalized by the thermal dipole cross section. (Left): At an early time for $\lambda=2$ shown for different angles, its angular average, the small-$b$ form \eqref{eq:smallbform}, and the approximated equilibrium expression \eqref{eq:iso_thermal_Cb}. (Right): At a small impact parameter $b\Teps=0.01$ as a function of time, normalized to equilibrium for $\lambda=2, \; 10$.
    }
    \label{fig:Cb}
\end{figure*}

{\textbf{Dipole cross section}.---}%
Jet quenching calculations or the gluon splitting rates often require the dipole cross section in impact parameter space, obtained via
\begin{align}
    C(\vb b) = \int\frac{\dd[2]{\vb q_\perp}}{(2\pi)^2}\left(1-e^{i\vb b\cdot\vb q_\perp}\right)C(\vb q_\perp).\label{eq:fouriertrafo}
\end{align}
For high-energetic particles, particularly the small $b \to 0$ form is important, which contains $\hat q$ and is known as the \emph{harmonic approximation}.
Unlike $\qhat$, we show that the small-$b$ form of $C(\vb b)$ is 
independent of a momentum cutoff $\Lambda_\perp$ and compare it with the full $C(\vb b)$. We start by splitting the integral \eqref{eq:fouriertrafo} into a small $q_\perp < \Lambda_\perp$ and large $q_\perp > \Lambda_\perp$ region. For a sufficiently high $\Lambda_\perp\gg T$, 
the high-$q_\perp$ integral becomes isotropic due to Eq.~\eqref{eq:limits_qperp} and depends on the medium distributions $f_a(\mbf k)$ only through $\mathcal N$ given by \eqref{eq:mathcal_N}.
With $\int_{\vb q_\perp}=\int\frac{\dd[2]{\vb q_\perp}}{(2\pi)^2}$, this leads to
\begin{align}
    \int_{\substack{\vb {q_\perp}\phantom{abcde}\\|\vb q_\perp|>\Lambda_\perp}}\!\!\!\!\!\!\!\!\!\!(1-e^{i\vb q_\perp\cdot \vb b})C(\vb q_\perp)=\frac{b^2 \CR g^4 \mathcal N}{8\pi}\log\frac{2 e^{1-\gamma_E}}{b\Lambda_\perp} +\mathcal O(b^4).
\end{align}
The small-$q_\perp$ part 
includes the jet quenching parameter 
\begin{align}
    \hat q^i(\Lambda_\perp,\tau)=\int_{\substack{\vb {q_\perp}\phantom{abcde}\\|\vb q_\perp|<\Lambda_\perp}} \!\!\!\!\!\!\!\!\!\!\!\!\!\!q_i^2 C(\vb q_\perp,\tau)\approx \hat q_0(\tau)\log\frac{\Lambda_\perp}{Q} + c_i(\tau). \label{eq:qhat}
\end{align}
Requiring Eq.~\eqref{eq:fouriertrafo} to be independent of $\Lambda_\perp$ for small $b$, forces an isotropic $\hat q_0(\tau)=\frac{C_R g^4 \mathcal N(\tau)}{4\pi}$.

The small-$b$ form of the kernel is then, in all generality,
\begin{align}
    C(\vb b,\tau)=\frac{1}{2}b_i^2 c_i(\tau)+\frac{b^2\hat q_0(\tau)}{2}\left(1-\gamma_E-\log\frac{bQ}{2}\right)+\mathcal O(b^4), \label{eq:smallbform}
\end{align}
with $b_i^2c_i(\tau)=b_y^2c_y(\tau)+b_z^2c_z(\tau)$.
This demonstrates that the knowledge of $\hat q(\Lambda_\perp)$ of the medium, with both $\hat q_0$ and $c_i$ in \eqref{eq:qhat}, suffices to accurately obtain the small distance form of the dipole cross section.

In our numerical calculations, we perform the integral \eqref{eq:fouriertrafo} on a momentum grid $q_\perp \in [q_\perp^{\mathrm{min}}, q_\perp^{\mathrm{max}}]$, $\phi \in [0, 2\pi]$, with sufficient angular and momentum resolution. To avoid discretization artifacts, we numerically extrapolate $q_\perp^{\mathrm{min}} \to 0$ and $q_\perp^{\mathrm{max}} \to \infty$ 
using the analytical limits in Eq.~\eqref{eq:limits_qperp}. The resulting cross section depends on $\vb b = b (\cos(\phi_b), \sin(\phi_b))$, with $\phi_b = 0$ ($\phi_b = \pi/2$) along (perpendicular to) the beam axis. 

The nonequilibrium dipole cross section is shown in \fig\ref{fig:Cb} for the angles $\phi_b = 0$ and $\pi/2$ as well as an angular averaged curve that reduces to $\langle C(\vb b) \rangle_{\phi_b} = \int_{\vb q_\perp}\left(1-e^{i\vb b\cdot\vb q_\perp}\right) \langle C(\vb q_\perp) \rangle_\phi$ and generally lies between the two angles. It is normalized to the equilibrium form $C_{\mathrm{eq}}(b)$, which we obtain by numerically evaluating the integrals with a thermal distribution for the same discretization and energy density $\varepsilon$. 
The left panel shows $C(\vb b, t)$ for an early time and $\lambda = 2$. 
We observe that it behaves qualitatively differently for small and large $b$ as compared to its thermal counterpart. 
At large distances, it significantly exceeds its equilibrium values, which confirms the higher efficiency of small-angle scatterings as we observed for $C(\vb \qperp)$. 
The small-$b$ region relevant for highly energetic partons (jet quenching) 
agrees well with the expression we derived in Eq.~\eqref{eq:smallbform} 
for the kernels in different directions and the angular averaged form.

The right panel of Fig.~\ref{fig:Cb} shows the time evolution of the dipole cross section at a small fixed distance $b\Teps=0.01$, normalized to equilibrium. We find that for both depicted couplings, the angular ordering of $C(\vb b)$ reverses after the star marker indicating high occupancy. 
It then remains at $C(\phi_b=0)>C(\phi_b=\pi/2)$, which aligns with the results found for the jet quenching parameter $\qhat^{i}$ in Ref.~\cite{Boguslavski:2023alu}. 
Remarkably. while for $\lambda=2$ the initial values exceed the equilibrium $C_{\mathrm{eq}}$, they stay below it for most of the time evolution. 
For the more realistic coupling $\lambda=10$, it remains smaller
during the entire evolution, leading to a reduction of jet quenching compared to thermal equilibrium. We note that this reduction is likely not sufficient to fully explain the negligibility of jet quenching for the pre-equilibrium medium claimed 
in Ref.~\cite{Andres:2019eus}, but it still shows
the relevance of nonequilibrium effects 
to genuinely describe the jet evolution.

Finally, we compare with common approximations of the dipole cross section. For the isotropic small-$q_\perp$ kernel \eqref{eq:analytic_collision_kernel}, the integral can be performed analytically,
\begin{align}
    C^{\mathrm{iso}}_{\mathrm{appr}}(b;T_\ast, m_D)&=\frac{C_Rg_s^2T_\ast}{2\pi}\left(\gamma_E+K_0(bm_D)+\log\frac{b m_D}{2}\right). \label{eq:iso_thermal_Cb}
\end{align}
For a general anisotropic distribution, this is an approximation of the full cross section. 
In Landau-matched thermal equilibrium, it reduces to $C^{\mathrm{eq}}_{\mathrm{appr}}=C^{\mathrm{iso}}_{\mathrm{appr}}(b;\Teps,m_D^{\mathrm{eq}})$.
Its ratio to the full 
thermal cross section is shown as a gray dash-dotted line in the left panel of Fig.~\ref{fig:Cb}. The approximated form exceeds it by more than $25\%$ for the small $b$ region relevant for jet quenching, signaling caution when using this form in practice.
\newline

{\textbf{Gluon splitting rate}.---}%
To study a consequence of the anisotropic collision kernel, we now consider the gluon splitting rate, which enters QCD kinetic theory simulations.
The rate for the splitting process $g\to gg$ with momenta $p\to zp + (1-z)p$, where $z$ is the energy fraction of the emitted gluon, can be calculated via \cite{Arnold:2002ja, Arnold:2002zm}
\begin{align}
    \gamma=\frac{p^4+p'^4+k'^4}{p^3p'^3k'^3}\frac{d_A\alpha_s}{2(2\pi)^3}\int_{\vb h}2\vb h\cdot \mathrm{Re} \vb F. \label{eq:gammarate}
\end{align}
The expression $\vb F$ is the solution to the integral equation
\begin{align} \label{eq:integral_eq_splitting}
    &2\vb h = i\delta E(\vb h)\vb F(\vb h)+\frac{g^2}{2}\int_{\vb q_\perp}C(\vb q_\perp)\\
    &\times \left[(3\vb F(\vb h)-\vb F(\vb h-p\vb q_\perp)-\vb F(\vb h-k\vb q_\perp)-\vb F(\vb h+p'\vb q_\perp)\right]\nonumber
\end{align}
with $\delta E(\vb h)=m_D^2/4\times (1/k+1/p-1/p')+h^2/(2pkp')$.
These rates are valid for all emitted gluon energies and were derived assuming an infinite medium and that the collision kernel $C(\vb q_\perp)$ does not change significantly during the formation time $\tform\sim\sqrt{\omega/\hat q}$ of a splitting process.
Regardless of these assumptions, they enter kinetic theory simulations by using the simplified isotropic forms \eqref{eq:analytic_collision_kernel} and \eqref{eq:iso_thermal_Cb} of the collision kernel \cite{Arnold:2002zm, AbraaoYork:2014hbk, Kurkela:2014tea, Kurkela:2015qoa, Kurkela:2018xxd, Du:2020zqg}.
Here, by generalizing Ref.~\cite{Aurenche:2002wq}, we apply a novel method developed in the companion paper \cite{Lindenbauer:preparation} (summarized in the Appendix) to solve the integral equation and evaluate Eq.~\eqref{eq:gammarate} for a general anisotropic $C(\vb q_\perp)$.

\begin{figure}
    \centering
    \includegraphics[width=\linewidth]{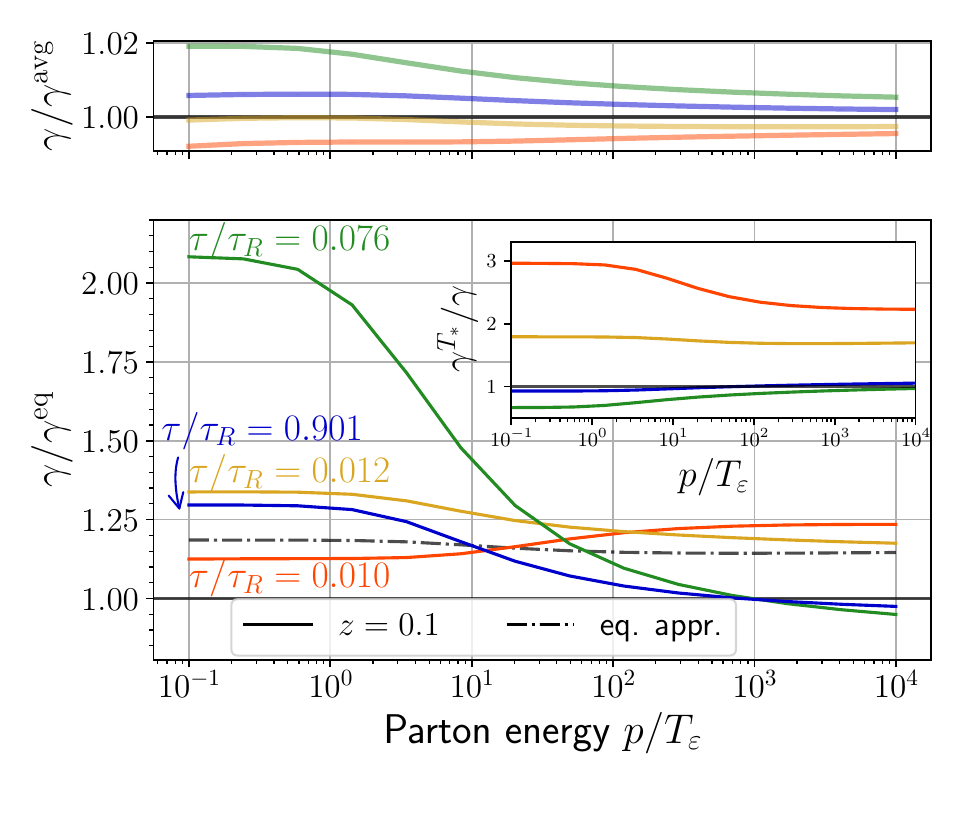}
    \includegraphics[width=\linewidth]{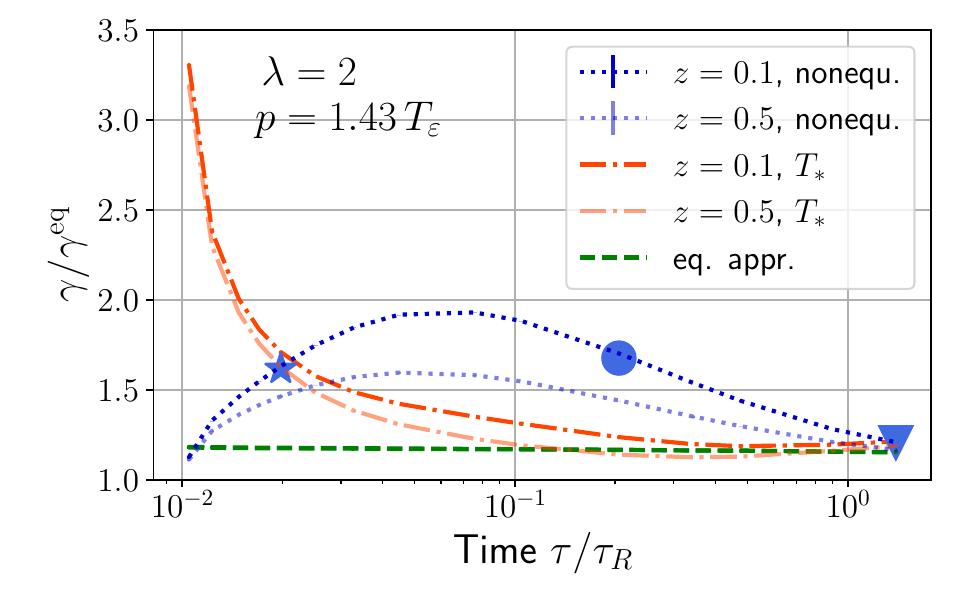}
    \caption{Splitting rate $\gamma$ from the anisotropic kernel $C(\vb q_\perp)$ in \eqref{eq:gammarate} for $\lambda=2$. (Top): 
    Normalized by the rate from the angular averaged kernel.
    (Center): Normalized by the equilibrium rate. Different times are color-coded. The black dash-dotted line shows the rate from $C^{\mathrm{eq}}_{\mathrm{appr}}$, 
    while the inset depicts the rate from the isotropically approximated kernel \eqref{eq:iso_thermal_Cb} over the nonequilibrium rate.
    (Bottom): Different rates for a fixed parton energy $p=1.43\Teps$ as functions of time. 
    }
    \label{fig:rate_ratio_nonequilibrium}
\end{figure}

Our results for the splitting rate $\gamma$ in the nonequilibrium medium for $\lambda=2$ 
are presented in Fig.~\ref{fig:rate_ratio_nonequilibrium}.
We first compare it to the splitting rate $\gamma^{\mathrm{avg}}$ from the angular averaged kernel $\langle C(b)\rangle_{\phi_b}$ in the upper panel. Notably, both quantities are seen to agree within just 2\% throughout the entire evolution. This motivates the applicability of an angular-averaged approximation of the full kernel for the splitting rate. 
We emphasize that this result concerns the collinear splitting rate $\gamma$ but may not be generalizable to other more differential observables \cite{Hauksson:2023tze, Barata:2024bqp}, for which the knowledge of the anisotropic $C(\vb q_\perp)$ may be relevant.

The central panel of Fig.~\ref{fig:rate_ratio_nonequilibrium} shows $\gamma$ normalized to the thermal rate for different times (color-coded) as a function of the parton energy $p$. We find that this ratio strongly depends on the parton energy and is enhanced by up to $100\%$ at low $p$. This far-from-equilibrium process indicates a particularly efficient branching of soft partons, which may lead to more soft gluons in the jetcone than anticipated using thermal rates.
We note that for large $p$, $\gamma$ behaves qualitatively differently, going from a light $25\%$ enhancement to close to the thermal rate. 

The central panel also shows that the rate obtained from $C^{\mathrm{eq}}_{\mathrm{appr}}$ (below \eqref{eq:iso_thermal_Cb}, black dash-dotted curve), that is often used as a toy model in calculations \cite{Arnold:2008zu, Andres:2020vxs}, overestimates the true thermal rate by over $20\%$ and thus must be taken with caution.
Using the approximation for the nonequilibrium kernel \eqref{eq:iso_thermal_Cb} commonly employed in QCD kinetic simulations to obtain $\gamma^{T_\ast}$ performs even worse, as shown in the inset. At the earliest times, $\gamma^{T_\ast}$ is three times as large as the nonequilibrium rate $\gamma$. This leads to an artificial enhancement of soft-gluon emissions and thus underestimates the formation time of the soft gluon bath during the thermalization process \cite{Baier:2000sb}. 

This is corroborated in the lower panel of Fig.~\ref{fig:rate_ratio_nonequilibrium}, which shows for a fixed parton energy $P=1.43\Teps$ the time evolution of the splitting rates for the nonequilibrium kernel (blue dotted), the isotropic approximated kernel $C^{\mathrm{iso}}_{\mathrm{appr}}$ in \eqref{eq:iso_thermal_Cb} (orange dash-dotted), and the equilibrium approximated kernel $C^{\mathrm{eq}}_{\mathrm{appr}}$, all normalized by the thermal rate. 
This confirms that $\gamma^{T_\ast}$ exhibits a qualitatively different behavior, which will impact 
QCD kinetic theory simulations, and thus, the description of the early stages in heavy-ion collisions.
\newline

{\textbf{Conclusion}.---}%
In this work, we studied jet momentum broadening beyond the harmonic approximation and the resulting gluon splitting rates for an anisotropic medium during the initial stages in heavy-ion collisions.

For the first time, we have determined the anisotropic full collision kernel $C(\vb q_\perp, \tau)$ of the nonequilibrium plasma in QCD kinetic theory, which measures the probability of jet broadening and characterizes the jet-medium potential.
The kernel has several advantages as compared to the jet quenching parameter $\qhat$, its second moment. 
Unlike $\qhat$, it avoids a spurious dependence on a transverse momentum cutoff 
and captures the entire broadening distribution, while $\hat q$ only characterizes the mean momentum transfer. 

For $C(\vb q_\perp, \tau)$, we find enhanced probability for soft scatterings
as compared to thermal equilibrium, corroborating
the use of the multiple-soft scattering approximation.
Interestingly, while its contribution to $\qhat$ is usually peaked at the Debye mass,
at early times and momentum transfer along the beam axis, the peak is shifted to lower momenta $q_\perp$. This suggests an angular-dependent effective screening scale, which is initially considerably lower than the Debye mass at small angles.
Although our kernel captures only a part of the plasma anisotropy due to the underlying isoHTL screening, our results provide a new practical avenue to go beyond isotropic screening prescriptions in future kinetic applications.

The kernel allows us to compute the jet-medium potential or dipole cross section $C(\vb b, \tau)$ in the impact parameter space.
To describe its small-distance behavior relevant for jet quenching, we derive and verify a simple cutoff-independent expression \eqref{eq:smallbform}.
We find that it 
is smaller than in a thermal system
for $\lambda=10$ during the entire pre-hydrodynamic evolution,
leading to a reduction of jet quenching during the initial stages. 
Although this remains a quantitatively modest effect, possibly explaining the 
relative suppression of jet quenching during the initial stages as proposed in \cite{Andres:2019eus},
our results 
highlight the importance of understanding nonequilibrium features in jet phenomenology, which is especially relevant for smaller collision systems.

To estimate the impact of the nonequilibrium kernel, we computed the gluon emission rates,
which are relevant for jet energy loss and used as input for QCD kinetic theory.
Remarkably, we find that using an angular-averaged collision kernel constitutes a good approximation for the nonequilibrium rate.
However, the latter significantly differs from the thermal rate, again demonstrating the importance of nonequilibrium dynamics.
We also observe that the approximated rate $\gamma^{T_\ast}$ differs drastically from the actual nonequilibrium $\gamma$, with differences of up to $300\,\%$ at early times.
Since it 
is employed in QCD kinetic theory simulations, our findings question the current treatment of inelastic splittings therein, with yet unexplored consequences.
It will be important to investigate this effect in future kinetic theory studies.

While we have focused here on the unpolarized gluon emission rate, several more differential observables are known to be sensitive to plasma and collision-kernel anisotropies \cite{Hauksson:2023tze, Hauksson:2023dwh, Barata:2024bqp}. Assessing the impact of our anisotropic kernel on these observables will be an important direction for future work.

\begin{acknowledgements}
    We wish to thank Néstor Armesto, Jo\~ao Barata, Andrey Sadofyev, and Sören Schlichting for insightful discussions and collaboration on related projects.
    FL is a recipient of a DOC Fellowship of the Austrian Academy of Sciences at TU Wien (project 27203). 
    This work is funded in part by the Austrian Science Fund (FWF) under Grant DOI 10.55776/P34455, and Grant DOI 10.55776/J4902 (FL).
    For the purpose of open access, the authors have applied a CC BY public copyright
    license to any Author Accepted Manuscript (AAM) version arising from this submission.
    The results in this paper have been achieved in part using the Austrian Scientific Computing (ASC) infrastructure, project 71444.
\end{acknowledgements}
\bibliography{bib}

\newpage
\appendix
\begin{figure}
    \centering
    \includegraphics[width=\linewidth]{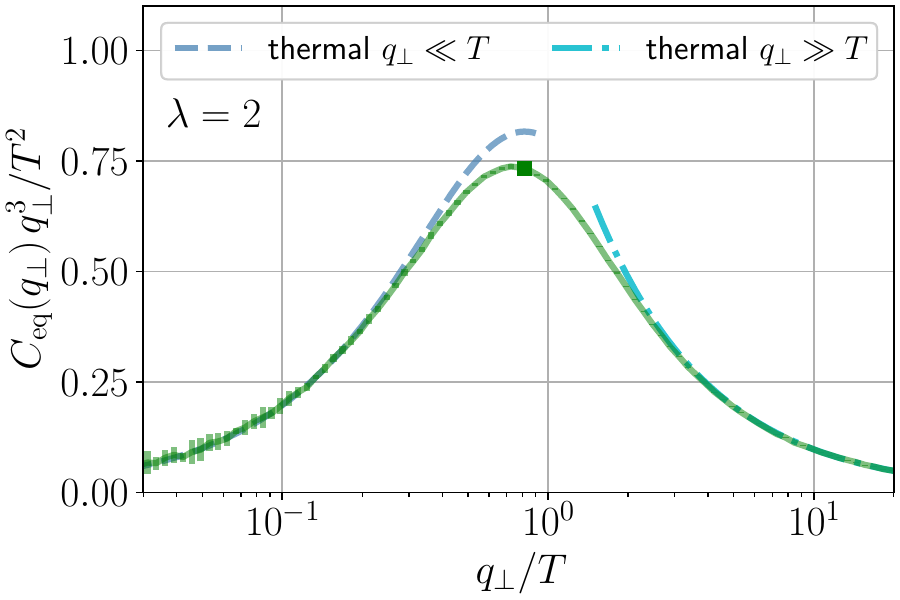}
    \caption{Collision kernel of a thermal system (green solid line) in comparison with its analytic estimates \eqref{eq:limits_general} for small and large cutoff. The box marks the Debye mass.
    }
    \label{fig:thermal}
\end{figure}
\begin{figure*}
    \centering
    \centerline{
    \includegraphics[width=0.33\linewidth]{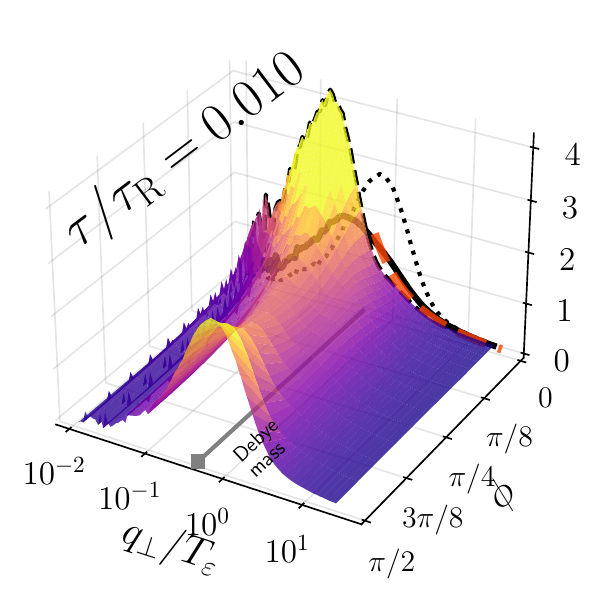}
    \includegraphics[width=0.33\linewidth]{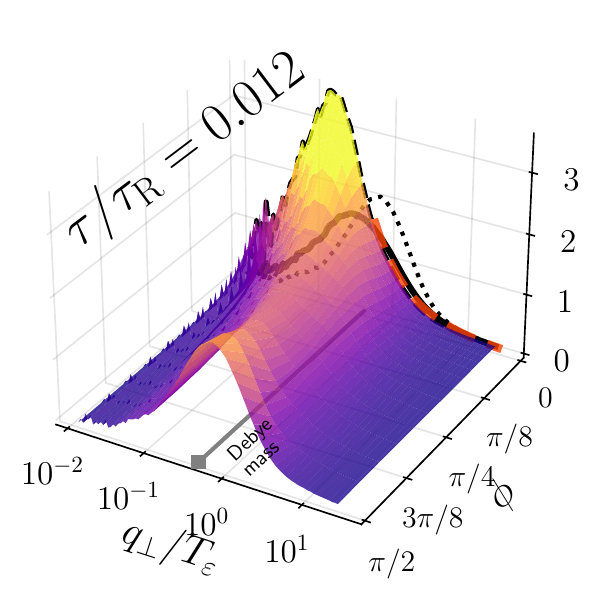}    
    \includegraphics[width=0.33\linewidth]{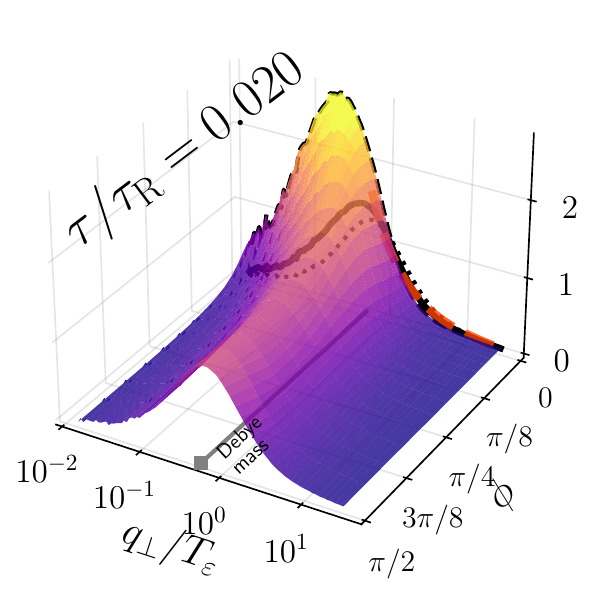}
    }
    \centerline{
    \includegraphics[width=0.33\linewidth]{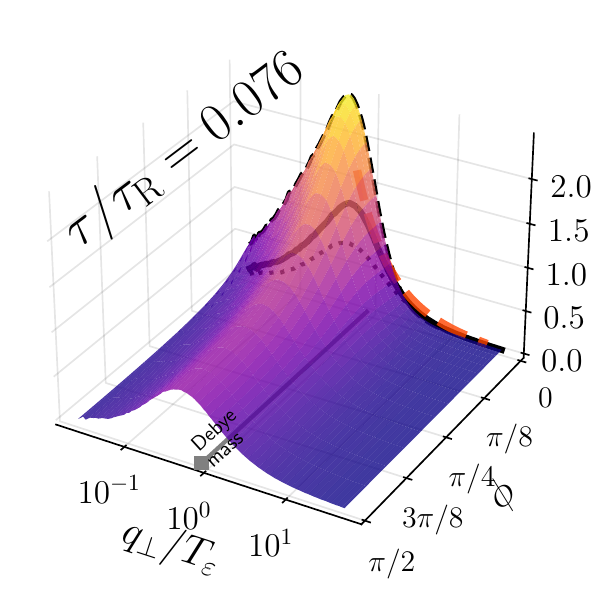}
    \includegraphics[width=0.33\linewidth]{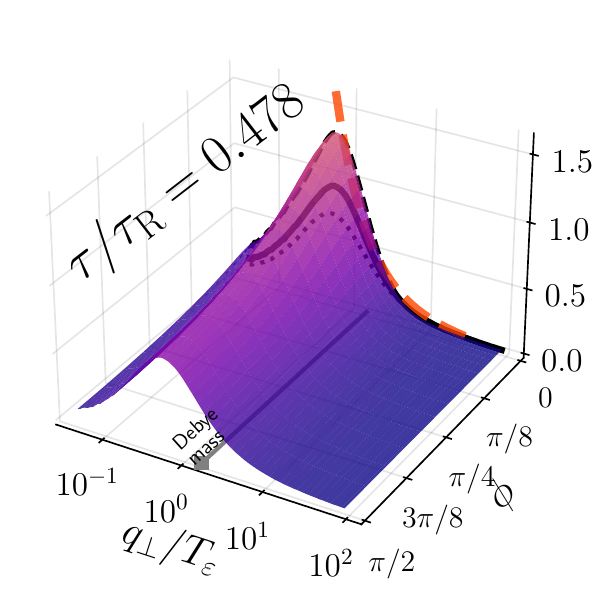}
    \includegraphics[width=0.33\linewidth]{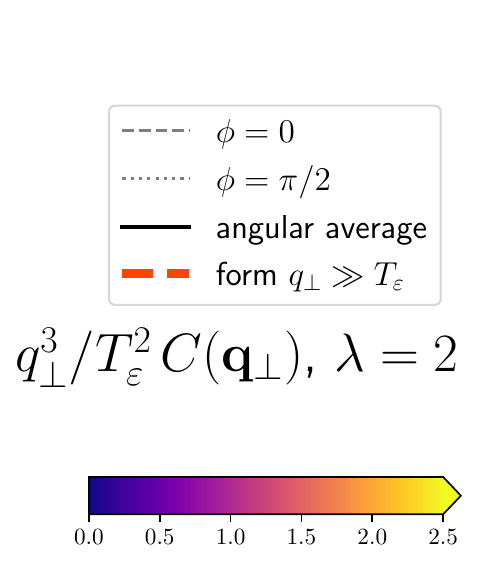}
    }
    \caption{Collision kernel $C(\vb q_\perp)$ of a gluonic plasma undergoing bottom-up thermalization for various times during the evolution for coupling $\lambda=2$. The back plane shows the projection of the angles $\phi=0$ and $\phi=\pi/2$, the angular average $\langle C(q_\perp)\rangle_\phi$ and the large $q_\perp$ result \eqref{eq:limits_qperp} in thermal equilibrium with the same energy density.
    }
    \label{fig:3dplots-lambda2}
\end{figure*}
\begin{figure*}
    \centering
    \centerline{
    \includegraphics[width=0.33\linewidth]{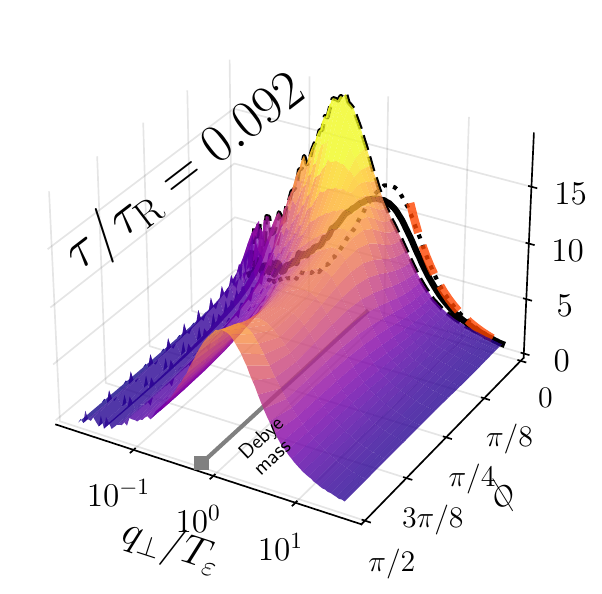}
    \includegraphics[width=0.33\linewidth]{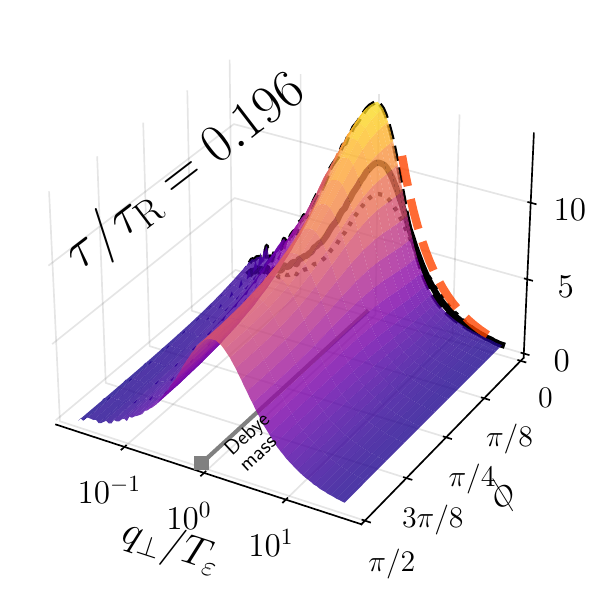}
    \includegraphics[width=0.33\linewidth]{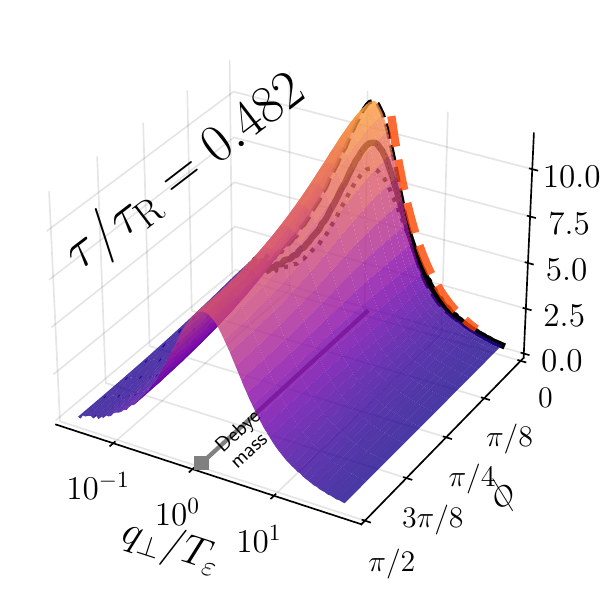}
    }
    \centerline{
    \includegraphics[width=0.33\linewidth]{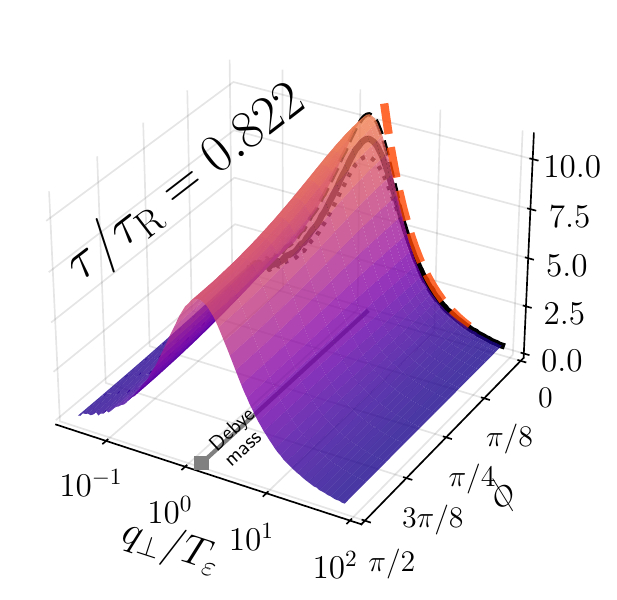}
    \includegraphics[width=0.33\linewidth]{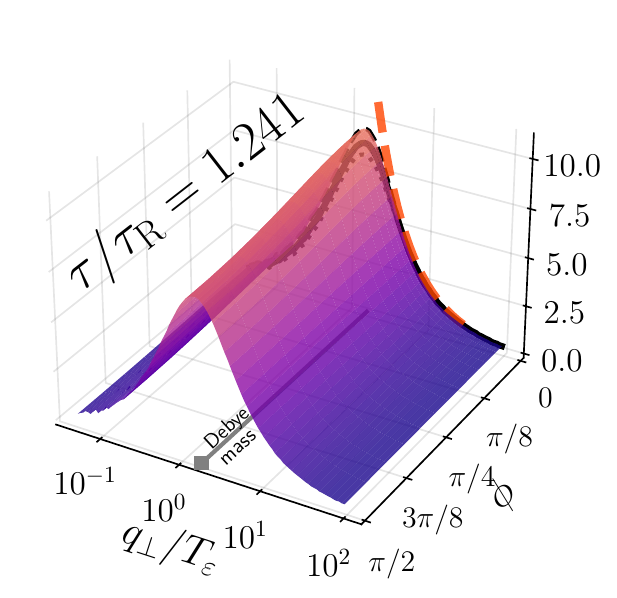}
    \includegraphics[width=0.33\linewidth]{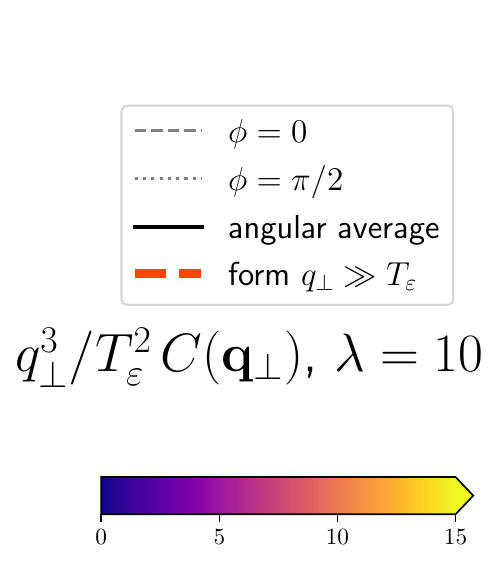}
    }
    \caption{
    We show the same collision kernel as in Fig.~\ref{fig:3dplots-lambda2}, but for the larger coupling $\lambda=10$.
    }
    \label{fig:3dplots-lambda10}
\end{figure*}

\section{Details on extracting the collision kernel}
Here, we provide details on the collision kernel and our implementation. We also discuss its symmetries and analytic limits.

\subsection{Collision kernel from the scattering rate}
Similar to Ref.~\cite{Boguslavski:2023waw} for the jet quenching parameter, we start with the elastic scattering rate
\begin{align}
    &\Gammael = \frac{1}{4p\nu_a}\int\frac{\dd[3]{\vb k}}{(2\pi)^32k}\frac{\dd[3]{\vb {p'}}}{(2\pi)^32p'}\frac{\dd[3]{\vb k'}}{(2\pi)^3 2k'}\,|\mathcal M|^2\\
    &\times(2\pi)^4\delta^4(P+K-P'-K')
     f(\vb k)(1\pm f(\vb k'))(1\pm f(\vb p')), \label{eq:scatteringrate}
\end{align}
from which the collision kernel can be obtained via \cite{Caron-Huot:2008zna}
\begin{align}
    C(\vb q_\perp)=(2\pi)^2\frac{\dd\Gammael}{\dd[2]{\vb q_\perp}}.
\end{align}
Note that Eq.~\eqref{eq:scatteringrate} is symmetric in $p'\leftrightarrow k'$ (switching the $u$ and $t$ channels), which allows us to enforce $p'>k'$ and produces a factor $2$.
We first integrate over $\vb k'$ using the delta function and then perform a variable substitution $\vb p'=\vb p+\vb q$. We pull out the integral over $\vb q_\perp$, and then arrive at Eq.~\eqref{eq:formula_Cq} from the main text,
\begin{align}
    C(\vb q_\perp)&=\frac{1}{2p\nu}\int\frac{\dd[3]{\vb k}}{(2\pi)^3}\dd {q_\parallel} \frac{|\mathcal M|^2}{8|\vb k||\vb k-\vb q||\vb p+\vb q|}\label{eq:formula_Cq_appendix}
    \\ &\times f(\vb k)(1+f(\vb k-\vb q))\delta(p+|\vb k|-|\vb p+\vb q|-|\vb k-\vb q|).\nonumber
\end{align}
We write this here for a plasma of gluons, which is also what we consider numerically.
In the limit $p\to\infty$, the matrix element for gluon-gluon scattering reduces to \cite{Boguslavski:2023waw}
\begin{align}
    \lim_{p\to\infty}\frac{|\mathcal M|^2}{p^2}=4 d_A C_A^2\left|G_R(P-P')_{\mu\nu}(P+P')^\mu(K+K')^\nu\right|^2,
\end{align}
where we use the full isotropic hard-thermal loop propagator $G_R$ to screen the internal soft gluon propagator due to medium effects as explained in Refs.~\cite{Arnold:2002zm, Boguslavski:2024kbd}.

The delta function can be rewritten as 
\begin{align}
        \delta(q_\parallel-k+k')&=\frac{k'}{kq}\delta\left(\cos\theta_{kq}-\frac{q_\parallel}{q}+\frac{t}{2kq}\right)\Theta(k-q_\parallel),
\end{align}
which only leads to a contribution if $k>\frac{q_\parallel+q}{2}$.
We then arrive at the expression as implemented in our numerical framework
\begin{align}
    C(\vb q_\perp)&=\frac{1}{16\nu(2\pi)^3}\int_0^{2\pi}\dd{\phi_{qk}}\int_{-\infty}^\infty\frac{\dd{q_\parallel}}{q}\int_{\frac{q+q_\parallel}{2}}^\infty\dd{k}\nonumber\\
    &\times\lim_{p\to\infty}\frac{|\mathcal M|^2}{p^2}f(\vb k)(1+f(\vb k-\vb q)). \label{eq:Cq_final}
\end{align}

\subsection{Symmetries of the collision kernel}
Let us discuss the symmetries of $C(\vb q_\perp)$ and $C(\vb b)$. First, in momentum space, we can write $C(\vb q_\perp)=C(q_z, q_y)$. Using the symmetry of the distribution function $f(k,\cos\theta_k)=f(k,-\cos\theta_k)$, which amounts to mirroring at the $z=0$ plane, we obtain $C(q_z,q_y)=C(-q_z,q_y)$. Additionally, the distribution function is symmetric under rotations in the transverse plane, implying $C(q_z,q_y)=C(q_z,-q_y)$. We thus obtain
\begin{align}
    C(q_z, q_y)=C(\pm q_z, \pm q_y). \label{eq:symmetry_Cq}
\end{align}
From the Fourier transform
\begin{align}
    C(b_z,b_y)&= \int\frac{\dd[2]{\vb q_\perp}}{(2\pi)^2}(1-e^{iq_z b_z+iq_yb_y})\,C(q_z,q_y)\\
    &=\int\frac{\dd[2]{\vb q_\perp}}{(2\pi)^2}(1-\cos(\vb q_\perp\cdot \vb b))\,C(q_z,q_y) \label{eq:fouriertrafo_real}
\end{align}
it follows immediately 
\begin{align}
    C(b_z,b_y)=C(\pm b_z,\pm b_y).
\end{align}
We can therefore map all angles $\phi_b > \pi/2$ into the first quadrant,
\begin{align}
    C(b,\phi_b)=\begin{cases}
        C(b,\phi_b), & 0 <\phi_b <\pi/2\\
        C(b,\pi-\phi_b), & \pi/2 < \phi_b < \pi\\
        C(b,\phi_b-\pi), & \pi < \phi_b <3/2 \pi\\
        C(b,2\pi-\phi_b), & 3/2 \pi < \phi_b < 2\pi
    \end{cases}\label{eq:Cb_symmetry}
\end{align}
Thus, evaluating a Fourier coefficient simplifies to
\begin{align}
    &\int_0^{2\pi}\dd{\phi_b}e^{-im\phi_b}D(b,\phi_b)\\
    &=\begin{cases}
        0, & m \text{ odd}\\
        4\int_0^{\pi/2}\dd\phi_b \cos m\phi_b \,D(b,\phi_b), & m\text{ even}
    \end{cases}\nonumber
\end{align}
for any function $D(\phi_b)$ that obeys the symmetry condition \eqref{eq:Cb_symmetry}.

\subsection{Analytic limits of the collision kernel}
Here, we discuss the limits of the collision kernel for large and small $q_\perp$.

First, we consider the limit of large transverse momentum transfer, i.e., $q_\perp \equiv |\vb q_\perp|\to\infty$. It is convenient to start with Eq.~\eqref{eq:formula_Cq_appendix}. For $p\to\infty$, $|\vb p+\vb q|\to p + q_\parallel$, when $\vb p=(0,0,|\vb p|)$, and then the delta function can be rewritten to constrain $q_\parallel$,
\begin{align}
    \delta(k-q_\parallel-|\vb k-\vb q|)\to \frac{q_\perp^2}{2(k-k_z)^2}\delta(q_\parallel-\frac{\vb q_\perp\cdot \vb k_\perp-q_\perp^2/2}{k-k_z}),
\end{align}
where in the normalization we have already taken the leading term in the limit $q_\perp\to\infty$. Thus, $q_\parallel\approx -q_\perp^2/(2(k-k_z))$, and $q\to\infty$. In this limit, screening effects can be neglected in the matrix element, and we may take the vacuum form
\begin{align}
    |\mathcal M|^2=-16 g^4d_A C_A^2\frac{su}{t^2}=16 g^4d_AC_A^2\frac{4p^2(k-k_z)^2}{q_\perp^4}.
\end{align}
Inserting this, we obtain
\begin{align}
    C(\vb q_\perp)
    &=\frac{4d_AC_A^2g^4}{\nu}\frac{1}{q_\perp^4}\int\frac{\dd[3]{\vb k}}{(2\pi)^3}\frac{k-k_z}{k}f(\vb k),
\end{align}
and the collision kernel is therefore proportional to the number density $J^0=n=\nu\int\frac{\dd[3]{\vb p}}{(2\pi)^3}f(\vb p)$ and the number current in the jet direction $J^z$. The number current vanishes when the distribution function is symmetric under $f(k,\cos\theta_k)=f(k,-\cos\theta_k)$. 
Using that, we obtain Eq.~\eqref{eq:limits_qperp} in the main text. More generally, including fermions, the large distance limit of the collision kernel is given by
\begin{align}
    \lim_{|\vb q_\perp|\to\infty}C(\vb q_\perp)\to\frac{g^4C_R\mathcal N}{q_\perp^4},
\end{align}
where $\mathcal N$
is given by Eq.~\eqref{eq:mathcal_N}.

We shall now consider the opposite limit, $|\vb q_\perp|\to 0$, where screening effects become important.
For isotropic systems, the $q_\parallel$ integral can be performed analytically using a sum rule \cite{Aurenche:2002pd} (see also Appendix B of Ref.~\cite{Boguslavski:2023waw}), which leads to Eq.~\eqref{eq:analytic_collision_kernel}.
However, for anisotropic distribution functions, this is more intricate, but still results in a similar behavior $1/q_\perp^2$, which we demonstrate in the following.
Starting with Eq.~\eqref{eq:Cq_final} for the collision kernel $C(\vb q_\perp)$,
we want to evaluate its small $q_\perp$ behavior, for which we need an explicit expression of the matrix element.
Using the notation of Ref.~\cite{Boguslavski:2023waw}, for the small $q_\perp$ region, it is sufficient to consider one of the dominant contributions to the matrix element that demonstrates the $1/q_\perp^2$ behavior, for instance, the term including the transverse propagator
\begin{align}
    |\mathcal M|^2&\ni
    \tilde c_2^2|G_T|^2.
    \label{eq:htl-matrix-el-general}
\end{align}
We can now use the sum rule \cite{Aurenche:2002pd} to perform the $q_\parallel$ integral analytically, which indeed leads to $\sim 1/q_\perp^2$ as in Eq.~\eqref{eq:limits_qperp} within the isoHTL approximation of the matrix element, even for an anisotropic distribution function.
Note that the prefactor can in general depend on the angle of $\vb q_\perp$.
However, a more careful analysis of its angular dependence is not necessary for our purposes, since we primarily use this limiting form to decrease the error in the numeric Fourier transform \eqref{eq:fouriertrafo}.

\subsection{The collision kernel in equilibrium}
We show our result for the collision kernel $C(\qperp)$ in thermal equilibrium in Fig.~\ref{fig:thermal}, where we also include
the analytic estimates \eqref{eq:limits_general} for both small and large $\qperp$.
We find excellent agreement with the analytic limits, which we take as a consistency check of our implementation.

\section{3D results}

To illustrate the continuous behavior of the collision kernel $C(\vb q_\perp, \tau)$ during the initial stages, we depict its form for several times in Fig.~\ref{fig:3dplots-lambda2} for $\lambda=2$ and in Fig.~\ref{fig:3dplots-lambda10} for $\lambda=10$. Both figures are qualitatively very similar. Initially, the kernel is anisotropic, with a large peak at $\phi=0$ and for values of $q_\perp$ below the Debye mass. A smaller peak at $\phi=\pi/2$  quickly dissolves during the evolution. At later times, the kernel becomes homogeneous in the angular direction and approaches the thermal form.

The back panels of all plots show the projection of the angle $\phi=0$ (dashed line) and $\phi=\pi/2$ (dotted line). Additionally, the angular average is shown as a thick black curve. One finds that all three curves converge towards the expected Eq.~\eqref{eq:limits_qperp} for large $q_\perp$, which is depicted as a thick dashed orange curve.

\section{Numerical method to solve the anisotropic rate equation}

As mentioned in the main text, we apply a novel method developed in the companion paper \cite{Lindenbauer:preparation} that extends Ref.~\cite{Aurenche:2002wq} to solve the self-consistent integral equation \eqref{eq:integral_eq_splitting}, and evaluate the rate $\gamma$ in \eqref{eq:gammarate} for a general anisotropic kernel $C(\vb q_\perp)$.
Equation \eqref{eq:integral_eq_splitting} is solved here in impact parameter space, where it can be written as
\begin{align}
    (A-D(z,\vb b)-B\nabla^2)\,\vb F(\vb b)=-2i\nabla \delta(\vb b), \label{eq:impactparameterspaceequation}
\end{align}
with $A=im_D^2/(4p) \,\times(1/z + 1/(1-z)-1)$, $B=i/(2pz(1-z))$ and $D(z,\vb b)=-\frac{1}{2} \left(C(\vb b) + C(z\vb b)+C((1-z)\vb b)\right)$, and $z$ is the energy fraction of the emitted gluon, i.e., $p\to zp + (1-z)p$.
Its numerical evaluation
uses the small-distance (Eq.~\eqref{eq:smallbform}) and large-distance limits 
\footnote{The integral \eqref{eq:fouriertrafo} is dominated when the exponent is $\mathcal O(1)$, where $\qperp\sim 1/b\to 0$, and $C(\qperp)\sim1/q_\perp^2$, leading to a logarithmic behavior for large $b$.}
of $C(\vb b)$, 
\begin{align}
    C(\vb b)\to\begin{cases}
        b^2(a_1\log b+a_2), & b\to 0\\
        a_3\log b + a_4, & b\to\infty,
    \end{cases}\label{eq:dipolecrosssection-limits}
\end{align}
where the coefficients $a_j$ may depend on the angle $\phi_b$.
We checked that our numerically computed $C(\vb b)$ obeys these analytic limits. 

The angular information is then decomposed 
in Fourier modes.
While in the isotropic case, $\vb F\sim \vb b$, and Eq.~\eqref{eq:impactparameterspaceequation} for small $b$ only has two linearly independent solutions, in general, there exist two linearly independent solutions for each Fourier mode, leading to infinitely many different solutions in total.
The boundary conditions are dictated for small $b$ by the delta function in Eq.~\eqref{eq:impactparameterspaceequation}, and by requiring that $F(\vb b)\to 0$ for $b\to \infty$, as well as that the rate \eqref{eq:gammarate} is finite.
While in the isotropic case, it is enough to solve two independent ordinary differential equations, in the anisotropic case, we need to solve $n_{\mathrm{fourier}}+3$ different systems of $n_{\mathrm{fourier}}$ coupled ordinary differential equations. 
We choose to solve for 7 and 11 Fourier modes to make sure that our results do not depend on the truncation of the Fourier series and refer to \cite{Lindenbauer:preparation} for more details on the numerical method.

\end{document}